\documentclass{ws-ijmpa}
\usepackage[super]{cite}
\usepackage{xcolor}
\usepackage{graphicx}
\usepackage[verbose,hypertexnames=false]{hyperref}
\hypersetup{colorlinks=false,allbordercolors=blue,pdfborderstyle={/S/U/W 1}}

\begin{document}

\markboth{Authors' Names}{Instructions for typing manuscripts (paper's title)}

%
\catchline{}{}{}{}{}
%

\title{Spacetime torsion signatures in neutrino oscillation physics}

\author{Capolupo Antonio}

\address{Dipartimento di Fisica E.R. Caianiello, Universita' di Salerno and INFN Gruppo Collegato di Salerno, Via Giovanni Paolo II, 132\\
Fisciano(SA), 84084, Italy}

\author{Monda Simone}

\address{Dipartimento di Fisica E.R. Caianiello, Universita' di Salerno and INFN Gruppo Collegato di Salerno, Via Giovanni Paolo II, 132\\
	Fisciano(SA), 84084, Italy}
	
\author{Pisacane Gabriele}

\address{Dipartimento di Fisica E.R. Caianiello, Universita' di Salerno and INFN Gruppo Collegato di Salerno, Via Giovanni Paolo II, 132\\
	Fisciano(SA), 84084, Italy}	

\author{Quaranta Aniello}

\address{School of Science and Technology, University of Camerino, Via Madonna delle Carceri\\
	Camerino, 62032, Italy}

\author{Serao Raoul}

\address{Dipartimento di Fisica E.R. Caianiello, Universita' di Salerno and INFN Gruppo Collegato di Salerno, Via Giovanni Paolo II, 132\\
	Fisciano(SA), 84084, Italy}	

\maketitle

\begin{history}
\received{Day Month Year}
\revised{Day Month Year}
\accepted{Day Month Year}
\published{Day Month Year}
\end{history}

\begin{abstract}
We report on recent results concerning neutrino oscillation in the presence of background torsion. In the context of Einstein-Cartan theory, we find new oscillation formulas for constant torsion and linearly time-dependent torsion. The oscillation formulas obtained depend on the orientation of the spin.
\end{abstract}


\ccode{PACS numbers: 03.65.$-$w, 04.62.+v}

\section{Introduction}
Recent discoveries in cosmology and astrophysics concerning the presence of dark energy \cite{DE2,DE5} and dark matter \cite{DM1,DM2,DM6,DM8} have motivated the study of alternative theories of gravitation beyond general relativity \cite{Capoz1,EX1,EX11,EX12}. These theories range from Brans and Dicke's scalar tensor model \cite{BransDicke} to modifying the Einstein-Hilbert action to models with torsion \cite{C-Dm-M-Q-S: 2024,Hehl,Shapiro,CUR1,CUR7,CUR10,NTorsion1,Cirillo-Lombardo}. On the other hand, studies on neutrinos and the effects of their presence in cosmology and astrophysics have been conducted in different directions. For example, neutrinos could be linked to baryonic symmetry and provide possible explanations for the dark matter of the universe \cite{ICE1,ICE2,ICE4,DUNE1,Curv0,Curv1,Kaplan,N1,N3,N5,N6,N7,N12,N17,NBSM1,NM2,KK,Pisacane,Luongo, Monda}. Neutrino oscillations have also been investigated from alternative and geometrical perspectives \cite{OLuongo,OLLuongo,Johns:2021,Capolupo:2016idi,Ahlu}. In this work, in the framework of QFT, we examine how neutrino propagation is modified when a background torsion field is present and analyse the resulting consequences for flavour oscillations. Here, we consider the Einstein-Cartan theory and analyze the case of constant torsion and a torsion field that varies linearly in time. We assume vanishing curvature. We shows that torsion introduces an additional energy splitting depending on the spin state, which modifies the oscillation formulas of quantum mechanics, quantum field theory, and CP asymmetry. 

It is important to emphasise that our treatment does not include the effective four-fermion interactions among neutrinos that would arise upon integrating out the non-propagating torsion in Eistein-Cartan theory \cite{Quaranta-Capolupo}. The torsion considered here is characterised as an external background field, which can be regarder as the mean torsion field produced by the spin density of generic fermions. 
\section{Dirac field and flavour mixing in the presence of torsion}

We investigate the impact of a background torsion, represented by the axial vector field $T^\mu(x)$, on Dirac fields in flat spacetime.  
In this case the Dirac equation are $	i \gamma^\mu \partial_\mu \Psi = m \Psi - \frac{3}{2} T_{\rho} \gamma^\rho \gamma^5 \Psi.$ By setting $u_{\vec{k}}^{r}(t)=e^{-iE^{r}t}u_{\vec{k}}^{r}$ and $v_{\vec{k}}^{r}(t)=e^{iE^{r}t}v_{\vec{k}}^{r}$, the standard plane-wave expansion of the Dirac field is recovered.  
In the scenario of constant torsion aligned along the third spatial axis, the solutions $u_{\vec{k}}^{r}$ and $v_{\vec{k}}^{r}$ are formally equivalent to those in flat spacetime, except that the effective mass becomes spin-dependent: $\widetilde{m}^{\pm} = m \pm \frac{3}{2} T^{3}$.  
Consequently, the torsion lifts the degeneracy between spin orientations, yielding the energy levels $E_{\vec{k}}^{\pm} = \sqrt{\vec{k}^{2} + (\widetilde{m}^{\pm})^{2}}\,.$ When the torsion field varies with time, the Dirac equation retains a similar structure. Let's use the ansatz

\begin{small}
	\begin{equation}
		\begin{array}{c}
			u_{\vec{k},\lambda}(t,\boldsymbol{x})=e^{i\boldsymbol{k}\cdot\boldsymbol{x}}
			\begin{pmatrix}
				f_{k}(t)\,\xi_{\lambda}(\hat{k})\\[1mm]
				g_{k}(t)\,\lambda\,\xi_{\lambda}(\hat{k})
			\end{pmatrix},\\[1mm]
			v_{\vec{k},\lambda}(t,\boldsymbol{x})=e^{i\boldsymbol{k}\cdot\boldsymbol{x}}
			\begin{pmatrix}
				g_{k}^{*}(t)\,\xi_{\lambda}(\hat{k})\\[1mm]
				-f_{k}^{*}(t)\,\lambda\,\xi_{\lambda}(\hat{k})
			\end{pmatrix},
		\end{array}
	\end{equation}
\end{small}

for positive and negative energy solutions.  
For a torsion field of the form $\breve{T}^{i} = \alpha^i t$, with $i=1,2,3$ and $\breve{T}^{0}$ constant in time, the time-dependent spinor components are

\begin{small}
	\begin{equation}
		\begin{cases}
			f_{\vec{k},\lambda}(t) = \exp\Big\{-i \frac{t^2}{2} \eta \lambda \alpha^i \hat{k}^i\Big\} \exp\{-i \omega_{k,\lambda} t\} C_{\vec{k},\lambda},\\[1mm]
			g_{\vec{k},\lambda}(t) = \frac{k + \eta \lambda \breve{T}^{0}}{\omega_{k,\lambda} + m} 
			\exp\Big\{-i \frac{t^2}{2} \eta \lambda \alpha^i \hat{k}^i\Big\} \exp\{-i \omega_{k,\lambda} t\} C_{\vec{k},\lambda},
		\end{cases}
	\end{equation}
\end{small}

where $\omega_{\vec{k},\lambda} = \sqrt{m^2 + (k + \eta \lambda \breve{T}^0)^2}$, $\eta$ denotes the coupling constant, $\lambda$ the helicity, $\vec{k}$ the momentum, and  $	C_{\vec{k},\lambda} = \frac{\omega_{k,\lambda}+m}{(2\pi)^{3/2} \sqrt{(\omega_{k,\lambda}+m)^2 + (k+\eta \lambda \breve{T}^0)^2}} \,.$

We denote the free Dirac fields in the torsion background as $\Psi_m^T \equiv (\nu_1, \nu_2, \nu_3)$.  
The flavor fields $\nu_\sigma^\alpha$ are obtained by the time-dependent unitary transformation $	\nu_\sigma^\alpha = J_\theta^{-1}(t) \, \nu_i^\alpha(x) \, J_\theta(t),$ where $(\sigma,i) = (e,1), (\mu,2), (\tau,3)$, and $J_\theta(t)$ is the mixing generator expressed as $	J_\theta(t) = J_{23}(t)\, J_{13}(t)\, J_{12}(t).$ This operator defines the plane-wave expansion of the flavor fields:

\begin{small}
	\begin{align}
		\nu_\sigma(x) = \sum_r \int \frac{d^3 \vec{k}}{(2\pi)^{3/2}} 
		\Big[ u_{\vec{k},i}^r \, \alpha_{\vec{k},\nu_\sigma}^r(t) + 
		v_{-\vec{k},i}^r \, \beta_{-\vec{k},\nu_\sigma}^{r\dagger}(t) \Big] 
		e^{i \vec{k} \cdot \vec{x}},
	\end{align}
\end{small}

where the flavor annihilation and creation operators are $	\alpha_{\vec{k},\nu_\sigma}^r(t) = J_\theta^{-1}(t) \, \alpha_{\vec{k},i}^r \, J_\theta(t)$ and $	\beta_{-\vec{k},\nu_\sigma}^{r\dagger}(t) = J_\theta^{-1}(t) \, \beta_{-\vec{k},i}^{r\dagger} \, J_\theta(t).$ These operators annihilate the flavor vacuum as usual: 
$\alpha_{\vec{k},\nu_\sigma}^r |0\rangle_f = \beta_{-\vec{k},\nu_\sigma}^r |0\rangle_f = 0$. For a momentum $\vec{k} = (0,0,|\vec{k}|)$, the flavor annihilation operators can be explicitly expressed in terms of the mass eigenstate operators:

\begin{align*}
	\alpha_{\vec{k},\nu_e}^r(t) &= c_{12} c_{13} \, \alpha_{\vec{k},1}^r + s_{12} c_{13} \Big[ (W_{12;\vec{k}}^{rr}(t))^* \alpha_{\vec{k},2}^r + \varepsilon^r (Z_{12;\vec{k}}^{rr}(t)) \beta_{-\vec{k},2}^{r\dagger} \Big],
\end{align*}

with $\varepsilon^r=(-1)^r$ and and corresponding expressions for the remaining flavors annichilators.  
To distinguish between the two cases, for constant torsion we set $W_{ij;\vec{k}}^{rr} = \Xi_{ij;\vec{k}}^{rr}\equiv u_{\vec{k},i}^{r\text{\ensuremath{\dagger}}}u_{\vec{k},j}^{s}=v_{-\vec{k},i}^{s\text{\ensuremath{\dagger}}}v_{-\vec{k},j}^{r}\,
$ and $
Z_{ij;\vec{k}}^{rr} = \chi_{ij;\vec{k}}^{rr}\equiv\varepsilon^{r}u_{\vec{k},1}^{r\text{\ensuremath{\dagger}}}
v_{-\vec{k},2}^{s}=-\varepsilon^{r}u_{\vec{k},2}^{r\text{\ensuremath{\dagger}}}v_{-\vec{k},1}^{s}\,$, while for time-dependent torsion we use $W_{ij;\vec{k}}^{rr} = \Pi_{ij;\vec{k}}^{rr}\equiv\left(u_{\vec{k},i}^{r},u_{\vec{k},j}^{s}\right)_{t}$ and $Z_{ij;\vec{k}}^{rr} = \Upsilon_{ij;\vec{k}}^{rr}\equiv\left(u_{\vec{k},i}^{r},v_{\vec{k},j}^{s}\right)_{t}$. Note that $W_{ij;\vec{k}}^{rs}=Z_{ij;\vec{k}}^{rs}=0$ for $r\neq s$. Explicitly, for costant torsion, one has

\begin{align}
	\Xi_{ij;\vec{k}}^{\pm\pm} &= N_i^\pm N_j^\pm \Bigg[1 + \frac{k^2}{(E_{\vec{k},i}^\pm + \widetilde{m}_i^\pm)(E_{\vec{k},j}^\pm + \widetilde{m}_j^\pm)}\Bigg] = \cos(\xi_{ij;\vec{k}}^{\pm\pm}),\\
	\chi_{ij;\vec{k}}^{\pm\pm} &= N_i^\pm N_j^\pm \Bigg[\frac{k_3}{E_{\vec{k},j}^\pm + \widetilde{m}_j^\pm} - \frac{k_3}{E_{\vec{k},i}^\pm + \widetilde{m}_i^\pm}\Bigg] = \sin(\xi_{ij;\vec{k}}^{\pm\pm}),
\end{align}

with $\widetilde{m}_i^\pm = m_i \pm \frac{3}{2} T^3$ and normalization factors $N_i^\pm = \sqrt{(E_{\vec{k},i}^\pm + \widetilde{m}_i^\pm)/(2 E_{\vec{k},i}^\pm)}$.  
The sign factor $\varepsilon^\pm = \mp 1$, and the energies satisfy $(E_{\vec{k},i}^\pm)^2 = \vec{k}^2 + (\widetilde{m}_i^\pm)^2$.  
The time evolution of the coefficients is governed by $	\Xi_{ij;\vec{k}}^{rs}(t)= |\Xi_{ij;\vec{k}}^{rs}| \, e^{i(E_{\vec{k},j}^{s}-E_{\vec{k},i}^{r}) t}$ and $\chi_{ij;\vec{k}}^{rs}(t) = |\chi_{ij;\vec{k}}^{rs}| \, e^{i(E_{\vec{k},j}^{s}+E_{\vec{k},i}^{r}) t}$. The canonical commutation relations are preserved through the unitarity condition: $\sum_r \big( |\Xi_{ij;\vec{k}}^{\pm r}|^2 + |\chi_{ij;\vec{k}}^{\pm r}|^2 \big) = 1, \quad i,j=1,2,3,\; j>i$.

For a time-dependent torsion field, the coefficients are denoted by $	\Pi_{ij;\vec{k}}^{rs}(t) = (u_{\vec{k},i}^r, u_{\vec{k},j}^s)_t$ and $	\Upsilon_{ij;\vec{k}}^{rs}(t) = (u_{\vec{k},i}^r, v_{\vec{k},j}^s)_t$. Explicitly, one finds:
\begin{align}
	\Pi_{ij;\vec{p}}^{ss}(t) &= (2\pi)^3 \, e^{-i (\omega_{p,s}^j - \omega_{p,s}^i) t} \, (C_{\vec{p},i}^s)^* (C_{\vec{p},j}^s) \Big[ 1 + \frac{|p + s \eta \breve{T}^0|^2}{(\omega_{p,s}^i + m_i)(\omega_{p,s}^j + m_j)} \Big],\\
	\Upsilon_{ij;\vec{p}}^{ss}(t) &= (2\pi)^3 \, e^{i t^2 \eta \alpha^i \hat{p}^i} \, e^{i (\omega_{p,s}^j + \omega_{p,s}^i) t} \, (C_{\vec{p},i}^s)^* (C_{\vec{p},j}^s)^* (p + s \eta \breve{T}^0) \Big[\frac{1}{\omega_{p,+}^j + m_j} - \frac{1}{\omega_{p,+}^i + m_i}\Big].
\end{align}

In the ultrarelativistic limit ($p \gg m_j$), these reduce to $\Pi_{\vec{p}}^{rr}(t) \rightarrow 1, \quad \Upsilon_{\vec{p}}^{rr}(t) \rightarrow 0,$
with the standard Minkowski-space coefficients.  
The canonical structure is preserved: $\sum_r \big( |\Pi_{ij;\vec{k}}^{\pm r}|^2 + |\Upsilon_{ij;\vec{k}}^{\pm r}|^2 \big) = 1$.

\section{Neutrino Oscillations in a Torsion Background}
\begin{figure}
	\centering
	\includegraphics[scale=0.5]{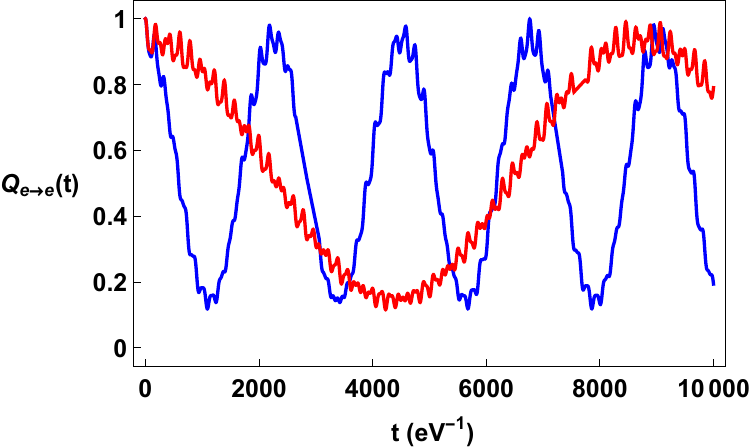}\includegraphics[scale=0.5]{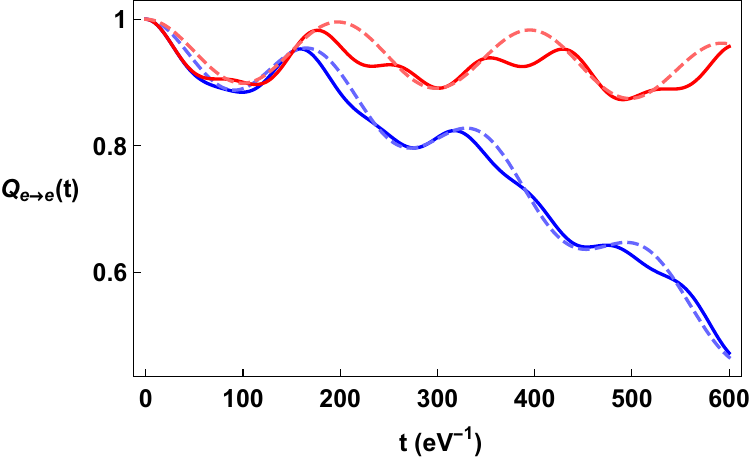}
	\caption{Transition probability $\mathcal{Q}^{\uparrow,\downarrow}_{\nu_e\to\nu_\tau}(t)$ with constant torsion. Left: spin-up (blue) and spin-down (red). Right: comparison with QM predictions (dashed). \label{fig:Pe-mu.3gen}}
\end{figure}

\begin{figure}
	\centering
	\includegraphics[scale=0.48]{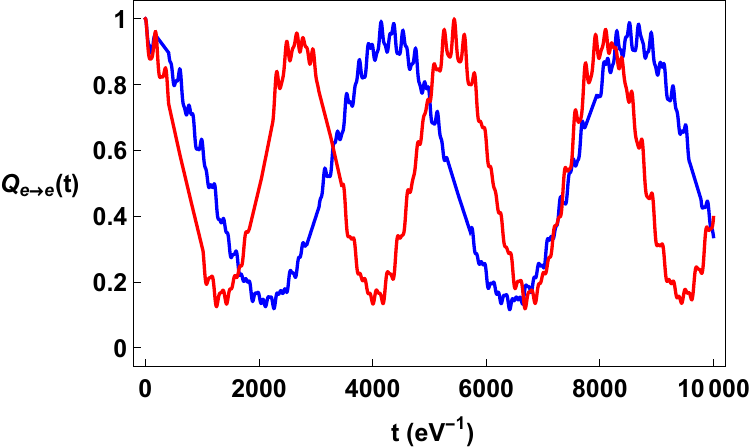} 
	\includegraphics[scale=0.48]{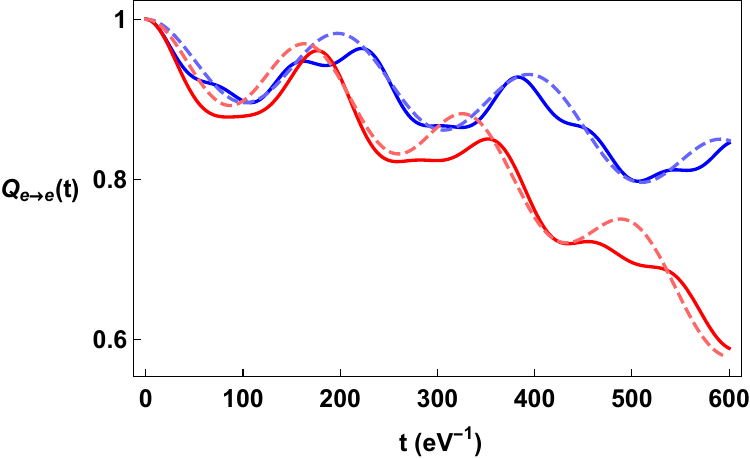}
	\caption{Transition probability for linearly time-dependent torsion. Left: spin-up (blue) and spin-down (red). Right: comparison with QM results (dashed). \label{fig:PeTauTime3Flavor}}
\end{figure}

We consider neutrinos propagating in costant and linearly time-dependent torsion background. By analyzing the flavor current \cite{N5} one derives the flavor charges:
\[
::\,Q_{\nu_{\sigma}}\,::=\sum_{r}\int d^{3}\boldsymbol{k}\,
\Big(\alpha_{\vec{k},\nu_{\sigma}}^{r\dagger}(t)\alpha_{\vec{k},\nu_{\sigma}}^{r}(t)-
\beta_{\vec{k},\nu_{\sigma}}^{r\dagger}(t)\beta_{\vec{k},\nu_{\sigma}}^{r}(t)\Big), \qquad \sigma=e,\mu,\tau,
\]
where $\left.::\cdots::\right.$, denoting the normal ordering with respect
to the flavor vacuum state $\left|0\right\rangle _{f}$.   
The oscillation formulas are obtained by evaluating, in the Heisenberg picture, the expectation value of $Q_{\nu_\sigma}$ on a neutrino flavor state:
$\left|\nu_{\vec{k},\sigma}^{r\dagger}(0)\right\rangle =\alpha_{\vec{k},\nu_\sigma}^{r\dagger}(0)\left|0\right\rangle _{f}$. At a fixed momentum $\vec{k}$ they are given by:
$
\mathcal{Q}_{\nu_{\rho}\rightarrow\nu_{\sigma}}^{r,\vec{k}}(t)  \equiv\left\langle \nu_{\vec{k},\rho}^{r}(t)\right|::\,Q_{\nu_{\sigma}}\,::\left|\nu_{\vec{k},\rho}^{r}(t)\right\rangle -{}_{f}\left\langle 0\right|::\,Q_{\nu_{\sigma}}\,::\left|0\right\rangle _{f}\,$. The oscillation formula:
\begin{align*}
	\mathcal{Q}_{\nu_{e}\rightarrow\nu_{e}}^{r,\vec{k}}(t) & =1-\sin^{2}(2\theta_{12})\cos^{4}(\theta_{13})\Big[|W_{12;\vec{k}}^{rr}|^{2}\sin^{2}(\Delta_{12;\vec{k}}^{r}t)
	+|Z_{12;\vec{k}}^{rr}|^{2}\sin^{2}(\Omega_{12;\vec{k}}^{r}t)\Big] \\
	& -\sin^{2}(2\theta_{13})\cos^{2}(\theta_{12})\Big[|W_{13;\vec{k}}^{rr}|^{2}\sin^{2}(\Delta_{13;\vec{k}}^{r}t)
	+|Z_{13;\vec{k}}^{rr}|^{2}\sin^{2}(\Omega_{13;\vec{k}}^{r}t)\Big] \\
	& -\sin^{2}(2\theta_{13})\sin^{2}(\theta_{12})\Big[|W_{23;\vec{k}}^{rr}|^{2}\sin^{2}(\Delta_{23;\vec{k}}^{r}t)
	+|Z_{23;\vec{k}}^{rr}|^{2}\sin^{2}(\Omega_{23;\vec{k}}^{r}t)\Big],
\end{align*}
with $r=\uparrow,\downarrow$, furthermore $\Delta_{ij;\vec{k}}^{r}=\frac{E_{j;\vec{k}}^{r}-E_{i;\vec{k}}^{r}}{2}$ and $\Omega_{ij;\vec{k}}^{r}=\frac{E_{j;\vec{k}}^{r}+E_{i;\vec{k}}^{r}}{2}$. Similar results are obtained for the transition between different flavors (see Ref.~\cite{C-Dm-M-Q-S: 2024}). Unitarity implies \(\mathcal{Q}_{\nu_{\rho}\to\nu_{e}}^{r,\vec{k}}(t)+\mathcal{Q}_{\nu_{\rho}\to\nu_{\mu}}^{r,\vec{k}}(t)+\mathcal{Q}_{\nu_{\rho}\to\nu_{\tau}}^{r,\vec{k}}(t)=1.\) These relations lead back to Pontecorvo's in the absence of torsion and in the ultrarelativistic limit $|\vec{k}|\gg m_{1},m_{2},m_{3}$. Then, the oscillation formulae are highly spin-dependent,  $\mathcal{Q^{\uparrow}}_{\nu_{\sigma}\rightarrow\nu_{\rho}}^{\vec{k}}(t)\neq
\mathcal{Q^{\downarrow}}_{\nu_{\sigma}\rightarrow\nu_{\rho}}^{\vec{k}}(t)$, since in QFT framework, the oscillation amplitudes and the frequencies
are spin depending. Notice that, in QM mixing treatment, the spin orientation affects only the frequencies $\Delta_{ij}$, being in this case:
$W_{ij;\vec{k}}^{\pm \pm} = 1$, $Z_{ij;\vec{k}}^{\pm \pm} =  \Omega_{ij;\vec{k}}^{\pm} = 0 $. To illustrate the impact of torsion we adopt representative parameter values: $m_{1}\approx10^{-3}\mathrm{eV}$, $m_{2}\approx9\times10^{-3}\mathrm{eV}$,
and $m_{3}\approx2\times10^{-2}\mathrm{eV}$ giving squared mass differences $\Delta m_{12}^{2}\approx7.56\times10^{-5}\,\mathrm{eV}^{2}$ and $\Delta m_{23}^{2}\approx2.5\times10^{-3}\,\mathrm{eV}^{2}$.  
The mixing angles are chosen as $\sin^{2}(2\theta_{13})=0.10$, $\sin^{2}(2\theta_{23})=0.97$, $\sin^{2}(2\theta_{12})=0.861$, with Dirac phase $\delta=\pi/4$. We also fix $k\simeq2\times10^{-2}\,\mathrm{eV}$ and torsion magnitude $|T^{3}|\simeq2\times10^{-4}\,\mathrm{eV}$.
Figures~\ref{fig:Pe-mu.3gen} and~\ref{fig:PeTauTime3Flavor} show the transition probabilities $\mathcal{Q}^{\uparrow,\downarrow}_{\nu_{e}\to\nu_{e}}(t)$ for constant and linearly time-dependent torsion. The right-hand panels compare the QFT predictions with the standard QM results.  
Spin dependence is manifest in both cases: the two spin states exhibit distinct oscillation patterns due to torsion-induced modifications of amplitudes and frequencies.

Quantum field theoretical corrections are most relevant at low momenta, where deviations from the QM approach are amplified. This remains true when torsion is included. Consequently, low-energy neutrino experiments such as PTOLEMY could potentially probe these effects, while high-energy facilities like DUNE~\cite{DUNE1} are expected to be largely insensitive.

The influence of torsion on $CP$ violation in neutrino oscillations, arising from the Dirac phase in the mixing matrix, is now considered. For fixed spin orientation, the $CP$ asymmetry is defined as \cite{N5}:
\[
\Delta_{\uparrow;CP}^{\rho\sigma}(t) \equiv \mathcal{Q}^{\uparrow}_{\nu_\rho\rightarrow\nu_\sigma}(t) + 
\mathcal{Q}^{\uparrow}_{\overline{\nu}_\rho\rightarrow\overline{\nu}_\sigma}(t), \quad \rho,\sigma = e,\mu,\tau,
\]
where the plus sign arises because antineutrino states carry a negative flavor charge. For the \(\nu_e \rightarrow \nu_\mu\) channel, the $CP$ asymmetry for spin \(r=\uparrow,\downarrow\) reads
\begin{align*}
	\Delta_{r;CP}^{e\mu}(t) &= 4 J_{CP} \Big[\,
	|W_{12;\vec{k}}^{\pm\pm}|^2 \sin(2\Delta_{12;\vec{k}}^\pm t) - 
	|Z_{12;\vec{k}}^{\pm\pm}|^2 \sin(2\Omega_{12;\vec{k}}^\pm t) \nonumber \\
	&+ (|W_{12;\vec{k}}^{\pm\pm}|^2 - |Z_{13;\vec{k}}^{\pm\pm}|^2) \sin(2\Delta_{23;\vec{k}}^\pm t) \nonumber \\
	&+ (|Z_{12;\vec{k}}^{\pm\pm}|^2 - |Z_{13;\vec{k}}^{\pm\pm}|^2) \sin(2\Omega_{23;\vec{k}}^\pm t) 
	- |W_{13;\vec{k}}^{\pm\pm}|^2 \sin(2\Delta_{13;\vec{k}}^\pm t) 
	+ |Z_{13;\vec{k}}^{\pm\pm}|^2 \sin(2\Omega_{13;\vec{k}}^\pm t)\, \Big]. 
	\label{eq:CPup}
\end{align*}
The spin-up coefficients correspond to \(W_{ij;\vec{k}}^{++}, Z_{ij;\vec{k}}^{++}\), while spin-down coefficients are \(W_{ij;\vec{k}}^{--}, Z_{ij;\vec{k}}^{--}\). Similarly, \(\Delta_{r;CP}^{e\tau}(t) = -\Delta_{r;CP}^{e\mu}(t)\).

Torsion induces a spin-dependent structure in the flavor vacuum, \(|0_f(t)\rangle\), resulting in distinct condensation densities for spin-up and spin-down components. These densities are obtained by evaluating expectation values of the number operators for free fields on the flavor vacuum: ${}_f\langle 0(t)| N_{\alpha_j,\vec{k}}^r |0(t)\rangle_f$ and ${}_f\langle 0(t)| N_{\beta_j,\vec{k}}^r |0(t)\rangle_f$ where $N_{\alpha_j,\vec{k}}^r = \alpha_{\vec{k},j}^{r\dagger}\alpha_{\vec{k},j}^r,$ $N_{\beta_j,\vec{k}}^r = \beta_{\vec{k},j}^{r\dagger}\beta_{\vec{k},j}^r,$ for $j=1,2,3.$
The condensate density may contribute to the dark sector of the Universe
\section{Conclusions}

In this study, we examined neutrino propagation within the Einstein-Cartan framework, incorporating torsion effects in a quantum field theory setting. We derived oscillation formulas that depend explicitly on neutrino spin, showing that torsion-induced energy splittings influence both oscillation frequencies and Bogoliubov amplitudes. Calculations were carried out in flat spacetime for both constant and linearly time-dependent torsion backgrounds. The torsion effects togheter with the QFT condensate effect are relevant at low neutrino momenta, suggesting that future low-energy experiments, such as PTOLEMY, could test the predictions presented here.

\section*{Acknowledgments}
Partial financial support from MIUR and INFN. is acknowledged. A. C. and G. L. also acknowledge the COST Action CA1511 Cosmology and Astrophysics Network for Theoretical Advances and Training Actions (CANTATA).

\section*{ORCID}

\noindent Capolupo Antonio - \url{https://orcid.org/0000-0002-8745-2522}

\noindent Monda Simone - \url{https://orcid.org/0009-0006-0387-8371}

\noindent Pisacane Gabriele - \url{https://orcid.org/0009-0006-6626-6655}

\noindent Quaranta Aniello - \url{https://orcid.org/0000-0002-8190-4989}

\noindent Serao Raoul - \url{https://orcid.org/0009-0001-2991-3036}

\end{document}